\documentclass[eqsecnum,showpacs,aps,twocolumn,nofootinbib,preprintnumbers]{revtex4}
\usepackage{graphicx,amssymb,bm,latexsym}
\newcommand{\bea}{\begin{eqnarray}}
\newcommand{\ena}{\end{eqnarray}}
\newcommand{\nn}{\nonumber\\}

\begin{document}

\preprint{KU-TP 003}
\preprint{WU-AP/256/06}
\preprint{hep-th/0607084}

\title{Supersymmetric Rotating Black Hole in a Compactified Spacetime}

\author{Kei-ichi Maeda\footnote{e-mail address: maeda@waseda.jp}$^{,a,b}$,
Nobuyoshi Ohta\footnote{e-mail address: ohtan@phys.kindai.ac.jp}$^{,c}$
and Makoto Tanabe\footnote{e-mail address: tanabe@gravity.phys.waseda.ac.jp}$^{,b}$}
\affiliation{$^a$DAMTP, Centre for Mathematical Sciences,
University of Cambridge, Wilberforce Road, Cambridge CB3 0WA, UK\\
$^b$Department of Physics Waseda University,
Okubo 3-4-1, Shinjuku-ku Tokyo 169-8555, Japan, \\
$^c$Department of Physics, Kinki University, Higashi-Osaka, Osaka 577-8502,
Japan
}

\date{\today}

\begin{abstract}
We construct a supersymmetric rotating black hole with asymptotically flat
four-dimensional spacetime times a circle, by superposing an infinite number
of BMPV black hole solutions at the same distance in one direction.
The near horizon structure is the same as that of the five-dimensional
BMPV black hole.
The rotation of this black hole can exceed the Kerr bound
in general relativity ($
q\equiv a/G_4 M=1$), if the size is
small.
\end{abstract}

\maketitle

\section{Introduction}

One of the most promising approaches to unification of fundamental
interactions is a superstring theory, or M-theory~\cite{string,Mtheory}.
Such unified theories are formulated in dimensions higher than four with
gravitational interactions. From this perspective,
it is important to study properties of black hole solutions in higher
dimensional theories~\cite{BH_higherD}, and it turns out that
they show a variety of new and interesting properties.
For example, there is no uniqueness theorem for
black holes in higher dimensions~\cite{Emparan_Reall,non_unique}.
In fact, we have a new type of black object, which is the so-called
black ring with horizon of a topology of $S^1\times S^2$.

Those solutions are obtained in the effective supergravity theories
in the low-energy limit. Among others, supersymmetric solutions are
expected to be important for the following reason.
The black hole solutions in the effective supergravity theories
will in general receive higher-order corrections
in the string coupling constant.
However, in the presence of unbroken partial supersymmetry, there are certain
properties on which such higher-order effects can be ignored.
For instance, the microstate counting of the black holes
is performed in the corresponding string solutions at the lowest order
in the string coupling, with result in perfect agreement with the
Bekenstein-Hawking entropy.
The underlying reason for this agreement is the remaining partial supersymmetry
because the numbers of dynamical degrees of freedom cannot change
under the change of the coupling constant in the BPS representation,
which guarantees that the countings of the states in the weak and strong
coupling regions should be the same.
Therefore, supersymmetric black hole (or black ring) solutions are
useful laboratories to examine properties which do not depend on perturbation
and are much discussed in the literature~\cite{BH_SUSY1,BH_SUSY2,BH_SUSY3}.

Since our world is four-dimensional, we should discuss four-dimensional
(or effectively four-dimensional) black holes with supersymmetry.
In higher dimensions, there is no uniqueness theorem and we can easily find
supersymmetric black holes for the case of spherically symmetric and
charged black holes.
On the other hand, in four-dimensional spacetime, it is known that
the Kerr-Newman black hole is the unique solution, but it is not
supersymmetric unless rotation vanishes.
Furthermore, there is even an argument that the event horizon of
a supersymmetric
black hole must be non-rotating~\cite{T}. It is thus interesting to
find rotating supersymmetric solutions which are (effectively) four-dimensional.
In five dimensions, there is no such restriction and it is already known
that there is a supersymmetric rotating
black hole called the BMPV black hole~\cite{BMPV},
and a supersymmetric black ring~\cite{EEMR04,Bena_Warner,EEMR05_1}.

Recently, Refs.~\cite{BD,EEMR05_2,GSY,BKW} constructed interesting rotating
black hole solutions with asymptotically flat four-dimensional spacetime.
Some of them use five-dimensional supersymmetric
black ring solutions in Taub-NUT base space~\cite{EEMR05_2,GSY,BKW}.
Here we present another example which is obtained by superposing
an infinite number of the BMPV black holes at the same distance
in one direction, which can be regarded as
a rotating supersymmetric
black hole in a compactified spacetime.
We discuss their properties including thermodynamic
quantities.
Similar techniques of compactification has been discussed
for extreme Reissner-Nordtsrom black holes in five dimensions~\cite{extreme_RN}.

For the case of non-BPS black holes, we can consider deformation of
a black hole in a compactified spacetime~\cite{FF}
and a transition from black string to black hole~\cite{G}.
The localized black hole may be preferred when the compactification
radius is large compared with the radius of the sphere of a black
string because of the Gregory-Laflamme instability ~\cite{GL}.
It is a subtle question if such a topology change occurs in the gravitational
theory.
In our case, however, because the solutions are supersymmetric,
there is no force between them and it is possible to superpose these
without any deformation. No such subtle instability is expected in our case,
which may be considered an advantage of our solutions.

This paper is organized as follows. In the next section, we present
our explicit solution for compactified BMPV black hole which is obtained
by compactification of one dimension. In Sec.~\ref{sec3}, we examine its
near horizon and asymptotic structures, entropy, rotation and supersymmetry.

\section{A compactified  BMPV solution}
\label{sec2}

In the previous paper~\cite{maeda_tanabe}, we showed that five-dimensional
rotating spacetimes can be constructed by compactifying 11-dimensional
stationary solutions with intersecting branes in M-theory.
We started with a generic form of the metric and solved the field
equations (the Einstein equations and the equations for form fields).
Assuming the intersection rule for branes~\cite{intersection_rule},
we derived the equations for each metric. We found that most of the metric
components are described by harmonic functions.
One metric component $f$, which describes a travelling wave,
is determined by the Poisson equation, whose source term is
given by the quadratic form of the ``gravi-electromagnetic" field ${\cal F}_{ij}$.
In some specific configuration of branes, e.g., for two intersecting
branes (M2$\perp$M5), the source term vanishes.
In these cases, all the harmonic functions are independent.
We can recover the BMPV rotating black hole solution in the simplest case.

Since the basic equations are linear (Laplace equations),
we can easily construct various solutions by superposing
those harmonics. Using this fact, here we construct an effectively
four-dimensional rotating black hole with supersymmetry.
In order to preserve 1/8 supersymmetry, we have to impose
that ${\cal F}_{ij}$ should be self-dual.

Let us start with the metric in five-dimensions:
\begin{eqnarray}
d\bar{s}_5^2=-\Xi^2
\left(dt+{\mathcal{A}_idx^i}
\right)^2+\Xi^{-1}ds_{\mathbb{E}^4}^2
\,,
\end{eqnarray}
where $\Xi=\left[H_2H_5(1+f)\right]^{-1/3}$.
$ds_{\mathbb{E}^4}^2$ is a four-dimensional Euclidean space
$(x^i)=(x,y,z,w)$.
$f$, $H_A (A=2, 5)$, and ${\cal A}_i$ are metric functions,
which will be determined later.

Adopting the hyperspherical coordinates:
\begin{eqnarray}
x+iy=\rho\cos\theta e^{i\phi},\quad
z+iw=-\rho\sin\theta e^{i\psi}
\,,
\end{eqnarray}
where $0\leq \phi, \psi <2\pi$ and $0\leq \theta \leq \pi/2$,
we have the line element of four-dimensional Euclidean space:
\begin{eqnarray}
ds_{\mathbb{E}^4}^2=d\rho^2+\rho^2\left(d\theta^2+ \cos^2\theta d\phi^2
+ \sin^2\theta d\psi^2\right)
\,.
\end{eqnarray}

The unknown metric functions $H_A (A=2,5)$, ${\cal A}_i$ and $f$ satisfy
the following equations:
\begin{eqnarray}
&&\partial^2 H_A=0,
\label{basic_eqs1}\\
&&\partial^2 f=0,
\label{basic_eqs2}\\
&&\partial^i \partial_{[i}{\cal A}_{j]}=0\,.
\label{basic_eqs3}
\end{eqnarray}
The simplest rotating spacetime solution  of these equations is given by
the so-called BMPV solution
\begin{eqnarray}
&&H_A=1+{Q_H^{(A)}\over \rho^2},\\
&&f={Q_0\over \rho^2},\\
&&{\cal A}_{\phi}={J_\phi \over 2\rho^2}\cos^2\theta,\\
&&{\cal A}_{\psi}={J_\psi \over 2\rho^2}\sin^2\theta
\,,
\end{eqnarray}
where $Q_H^{(A)}$, $Q_0$, $J_\phi $, and $J_\phi$ are charges and angular
momenta. Note that supersymmetry of the BMPV solution imposes the condition
$J_\phi=-J_\psi$. The electric fields are given by
\begin{eqnarray}
E_j^{(A)}=\partial_j\left({1\over H_A}\right)
\,.
\end{eqnarray}
Since the origin of coordinate system can be shifted by any distance,
we can move such a black hole to any position.
Also because the basic equations (\ref{basic_eqs1}), (\ref{basic_eqs2}),
and (\ref{basic_eqs3}) are linear, we can superpose those black hole
solutions at different positions.
Especially, if we superpose an infinite number of black holes, each of
which is separated by the same distance in one direction, we obtain
a periodic BMPV black hole solution.
It can be regarded as a deformed BMPV black hole in a compactified
five-dimensional spacetime.
We call it a compactified BMPV black hole (a CBMPV black hole).

Let us now present its explicit solution.
Suppose that an infinite number of black holes
exist along the $w$-axis with the same coordinate distance $2\pi R_5$,
with $R_5$ denoting the compactification radius at infinity.
By superposing those black hole solutions, we have
\begin{eqnarray}
H_A&=& 1+Q_H^{(A)} \sum_{n=-\infty}^{\infty} \frac{1}{r^2+(w+2\pi n R_5)^2},
\\
f&=& Q_0 \sum_{n=-\infty}^{\infty} \frac{1}{r^2+(w+2\pi n R_5)^2},
\\
{\cal A}_{\phi}&=& {J_\phi\over 2}
\sum_{n=-\infty}^{\infty} \frac{x^2+y^2}{[r^2+(w+ 2\pi n R_5)^2]^2},
\\
{\cal A}_{\psi}&=& {J_\psi\over 2}
\sum_{n=-\infty}^{\infty}\frac{z^2+(w+ 2\pi n R_5)^2}{[r^2+(w+ 2\pi n R_5)^2]^2}
\,,
\end{eqnarray}
where $r^2\equiv x^2+y^2+z^2$.

\begin{widetext}
~
Introducing a new function
\begin{eqnarray}
F(\xi,\eta)\equiv \sum_{n=-\infty}^{\infty} \frac{1}{\xi^2+(\eta+2\pi n )^2}
= {\sinh \xi \over 2\xi \left(
\cosh \xi-\cos \eta \right)}
\,,
\label{f}
\end{eqnarray}
and setting $\bar{r}= r/ R_5$ and $\bar{w}= w/R_5$, we obtain
\begin{eqnarray}
H_A&=&
1+{Q_H^{(A)}\over 2  R_5^2}F(\bar{r},\bar{w}) =
1+{Q_H^{(A)}\over 2  R_5^2}
{\sinh \bar{r}\over \bar{r}\left(
\cosh \bar{r}-\cos \bar{w} \right)},
\label{ha}
\\
f&=&
{Q_0\over 2  R_5^2} F(\bar{r},\bar{w}) =
{Q_0\over 2  R_5^2}{\sinh \bar{r}\over
 \bar{r}\left(
\cosh \bar{r}-\cos \bar{w} \right)},
\\
{\cal A}_{\phi}&=&
-{J_\phi\over 2R_5^2}
{\bar{x}^2+\bar{y}^2\over 2\bar{r}}{\partial
F(\bar{r},\bar{w}) \over \partial \bar{r}}
=
{J_\phi\over 8R_5^2}
{\left(\bar{x}^2+\bar{y}^2\right)\over \bar{r}^2}
\left[
{\left(\cosh \bar{r}\cos \bar{w}-1
\right)\over \left(\cosh \bar{r}-\cos \bar{w}\right)^2}
+{\sinh \bar{r}\over \bar{r}
\left(\cosh \bar{r}-\cos \bar{w}\right)}
\right],
\\
{\cal A}_{\psi}&=&
{J_\psi\over 2R_5^2}
\left(F+{\bar{x}^2+\bar{y}^2\over 2\bar{r}}{\partial
F(\bar{r},\bar{w})\over \partial \bar{r}}\right)
\nn
&=&
{J_\psi\over 8R_5^2}
\left[
-{(\bar{x}^2+\bar{y}^2)\over \bar{r}^2}
{\left(\cosh \bar{r}\cos \bar{w}-1
\right)\over \left(\cosh \bar{r}-\cos \bar{w}\right)^2}
+{\left(\bar{r}^2+\bar{z}^2 \right)\over \bar{r}^2}
{\sinh \bar{r}
\over  \bar{r}
\left(\cosh \bar{r}-\cos \bar{w}\right)}
\right]
\,.
\label{apsi}
\end{eqnarray}

\end{widetext}

Since this solution is periodic in the $w$-direction with the period
$2\pi R_5$, it can be regarded as a deformed BMPV black hole
in a compactified five-dimensional spacetime ($-\pi R_5 \leq w \leq \pi R_5$)
with supersymmetry (if $J_\phi=-J_\psi$).
In the following sections, we analyze the properties of this solution.

\section{Spacetime structure of a compactified BMPV black hole}
\label{sec3}

\subsection{horizon}

The horizon exists at $(r, w)=(0, 0)$.
Setting $\bar{x}=\epsilon\cos\theta\cos\phi$,
$\bar{y}=\epsilon\cos\theta\sin\phi$,
$\bar{z}=\epsilon\sin\theta\cos\psi$,
$\bar{w}=\epsilon\sin\theta\sin\psi$
($\epsilon \ll 1$),
we find the behavior of the metrics near horizon to be
\bea
&&H_A=1+{Q_H^{(A)}\over   R_5^2\epsilon^2},~~~~~~
f={Q_0\over   R_5^2\epsilon^2}, \\
&&A_\phi={J_\phi\over  2 R_5^2\epsilon^2}\cos^2\theta,~~
A_\psi={J_\psi\over  2 R_5^2\epsilon^2}\sin^2\theta.
\ena
This is exactly the same near-horizon structure as
that of the BMPV black hole.
Hence the horizon structure is not modified
by superposion of an infinite number of black holes.
This is due to the BPS properties of the solutions which guarantee the
no-force condition between the black holes of the solutions.
It is just the same as the case of superposition of
extreme Reissner-Nordstrom black holes.

The area of the horizon is given by
\bea
A_H=2\pi^2\sqrt{Q_0 Q_H^{(2)} Q_H^{(5)}-J^2/8}\,,
\ena
where
$J^2=(J_\phi^2+J_\psi^2)/2$.

\subsection{asymptotic structure}

\begin{figure}[h]
\includegraphics[scale=1]{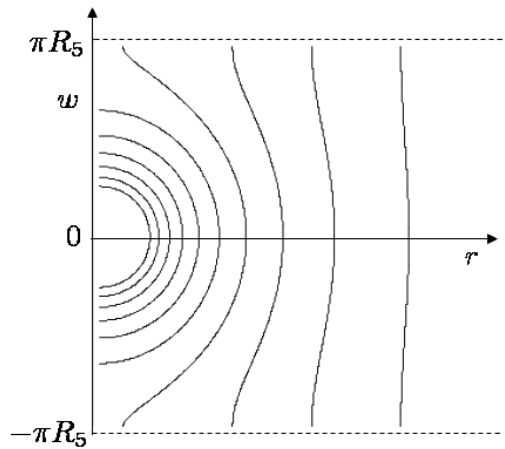}
\caption{Equipotential lines for
``gravitational potential" $f(r,w)$.
The dotted lines ($w= - \pi R_5$ and $w=\pi R_5$) are identified
to compactify the $w$-direction.}
\label{fig0}
\end{figure}

The metric describes the five-dimensional spacetime, but it behaves effectively
as the four-dimensional spacetime (times a circle) when an observer is far from
the black hole. In fact, $w$-direction is compactified with the period
$2\pi R_5$, while the other spatial directions are not ($0\leq r<\infty$).
In Fig.~\ref{fig0}, we depict equipotential lines for
``gravitational potential" $f$ (or $H_A$).
The origin ($(r,w)=(0,0)$) is horizon of a black hole.
The potential depends nontrivially on the compact direction $w$
near a black hole, but the metric asymptotically approaches flat
four-dimensional Minkowski times the $w$-circle. Near infinity, the
metric becomes independent of the coordinate $w$.

\begin{figure}[h]
\includegraphics[scale=.9]{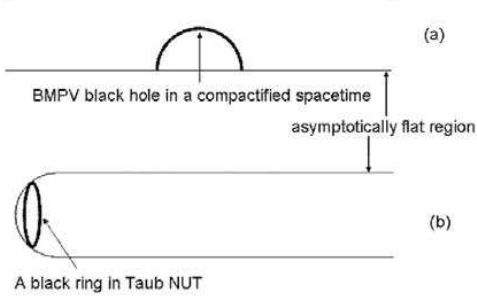}
\caption{The schematic pictures of a CBMPV black hole (a)
and  a RING black hole (b).}
\label{fig2}
\end{figure}

The structure of this black hole is schematically
given in Fig.~\ref{fig2}~(a).
For reference, we also show the structure of
another supersymmetric  rotating black hole solution with
asymptotically flat four-dimensional spacetime,
which is obtained by use of a
black ring solution in Taub-NUT base space~\cite{EEMR05_2,GSY,BKW}.
We call it a RING black hole.

The RING black hole is embedded in a curved Taub-NUT space. The asymptotic
region is four-dimensional Minkowski spacetime (Fig. \ref{fig2}~(b)).
The CBMPV  black hole exists in a compactified
5D spacetime  obtained by compactification of
the fifth direction (Fig. \ref{fig2}~(a)).
Both spacetimes have asymptotically flat 4D Minkowski spacetime with
small compactified dimension (its radius is $R_5$),
but those horizons are five-dimensional.

Let us now examine how the asymptotic four-dimensional spacetime looks like.
Introducing the three-dimensional spherical coordinates
($\bar{r},\Theta,\Phi$),
which are defined by the transformations
\bea
&&
\bar{x}=\bar{r} \sin \Theta \cos \Phi,
\nn
&&
\bar{y}=\bar{r} \sin \Theta \sin \Phi,
\nn
&&
\bar{z}=\bar{r} \cos \Theta\,,
\ena
our metric is rewritten as
\begin{widetext}
\bea
d\bar{s}^2&=&-\Xi^2\left[
d\bar{t}+\bar{\cal A}_\phi d\Phi
+{\bar{\cal A}_\psi\over (\bar{r}^2\cos^2\Theta+\bar{w}^2)}
(-\bar{w}\cos\Theta d\bar{r}+\bar{r} \bar{w}
 \sin\Theta d\Theta+\bar{r}\cos\Theta d\bar{w})
\right]^2
\nn
&&~~~
+\Xi^{-1}\left(d\bar{r}^2+\bar{r}^2d\Omega^2+d\bar{w}^2
\right),
\ena
where
\bea
&&
\bar{\cal A}_\phi={J_\phi\over 8 R_5^3}
\sin^2\Theta\left[{\sinh \bar{r}\over
\bar{r}(\cosh\bar{r}-\cos \bar{w})}
+{\cosh\bar{r}\cos \bar{w}-1\over (\cosh\bar{r}-\cos \bar{w})^2}\right],
\nn
&&
\bar{\cal A}_\psi={J_\psi\over 8 R_5^3}
\left[(1+\cos^2\Theta){\sinh \bar{r}\over
 \bar{r}(\cosh\bar{r}-\cos \bar{w})}
-\sin^2\Theta{\cosh\bar{r}\cos \bar{w}
-1\over (\cosh\bar{r}-\cos \bar{w})^2}\right]
\,.
\ena
\end{widetext}

In the asymptotic region ($\bar{r}\gg \bar{w}$),
we can make a Kaluza-Klein reduction of the near-asymptotic metric
in the $w$ direction.
We then find the effective four-dimensional metric, which can be written
in the Einstein frame as
\bea
d\bar{s}_4^2&=&-\Xi^{3/2}\left(
d\bar{t}+\bar{\cal A}_\phi d\Phi
\right)^2
\nn
&+&\Xi^{-3/2}\left[d\bar{r}^2+\bar{r}^2\left(d\Theta^2+
\sin^2\Theta d\Phi^2\right)
\right]\,,
\ena
where
\bea
&&
H_A=1+{1\over 2R_5^2}{Q_H^{(A)}\over \bar{r}},
\nn
&&
f={1\over 2 R_5^2}{Q_0\over \bar{r}},
\nn
&&
\bar{\cal A}_\phi= { 1 \over 8 R_5^3} {J_\phi\over \bar{r}}\sin^2\Theta
\,.
\ena

Comparing this with the asymptotic form of the four-dimensional Kerr-Newman
metric, we get a gravitational mass of the black hole $M$ and a rotation
parameter $a$, which is the angular momentum per unit mass:
\bea
&&G_4 M={  (Q_0+Q_H^{(2)}+Q_H^{(5)})\over 8  R_5}=
{G_5 M_{\rm ADM}\over 2\pi R_5},
\nn
&&G_4 Ma={ J_\phi\over 16R_5}
\,,
\ena
where $G_4=G_5/(2\pi R_5)$ is the four-dimensional
gravitational constant and $M_{\rm ADM}$ is
the ADM mass of the BMPV black hole.
The rotation parameter  is given by
$a=J_\phi/[2(Q_0+Q_H^{(2)}+Q_H^{(5)})]$.

{}From the asymptotic form of the electric fields
\bea
E_r^{(A)} \sim  {1\over 2  R_5} {Q_H^{(A)}\over r},
\ena
we also find the charges in four-dimensions are
\bea
Q_A =  {Q_H^{(A)}\over 2 R_5}.
\ena

\subsection{entropy}

The entropy of the present black hole is given by that of
the BMPV black hole, i.e.
\bea
&&S_{\rm CBMPV}={A_H\over 4G_5}=
{\pi^2\over 2G_5}\sqrt{Q_0 Q_H^{(2)} Q_H^{(5)}-{J^2\over 8}}.
~~~~
\ena
Setting $Q^{(A)}=\alpha^{(A)}Q_0 ~(A=2,5)$,
and assuming supersymmetry, i.e. $J_\phi=-J_\psi(=J)$,
we find
\bea
S_{\rm CBMPV}={\pi r_g^2\over 2G_4}\sqrt{{1\over 2}\left({1\over \lambda^2}-
q^2\right)},
\ena
where $r_g\equiv 2G_4 M$ is the Schwarzschild radius
of a black hole mass $M$, and
\bea
\lambda^2&\equiv &{ (1+\alpha^{(2)}+\alpha^{(5)})^3
 \over 32\alpha^{(2)}\alpha^{(5)}}\times {r_g\over  R_5},~~\nn
&& \hspace{-5mm}
({\rm ratio~of~the~BH~size~to~the~compactification~scale})
\nn
q^2 &\equiv &{a^2\over G_4^2M^2},~~
({\rm Kerr~rotation~parameter})
\nn
e_{A} &\equiv &{Q_A\over G_4M}={4\alpha^{(A)}\over 1+\alpha^{(2)}
+\alpha^{(5)}},~~
(A=2,5)\nn
&&
({\rm normalized~charges})
\ena

For reference, we show the entropy of Kerr-Newman black hole, which is given by
\bea
S_{\rm KN}={\pi r_g^2\over 2 G_4}\left[1-{e^2\over 2}+\sqrt{1-
q^2-e^2}\right]\,,
\label{KN}
\ena
where $r_g$ and $q$ are the same as above and $e$ is a normalized charge
of Kerr-Newman black hole.

\subsection{rotation}

For the Kerr-Newman black hole,
we have one constraint for the existence of regular horizon,
which is $q^2+e^2\leq 1$. The equality
gives the extreme limit. So we find
maximal rotation from this condition, that is
$q^2=1-e^2$. The Kerr black hole ($e=0$) gives
maximal rotation $q=1$ in the extreme limit.

\begin{figure}[h]
\includegraphics[scale=1]{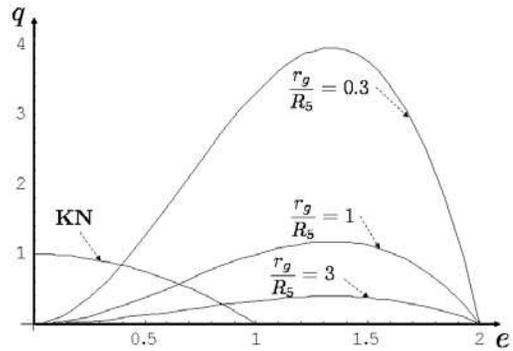}
\caption{A rotation parameter $q$ with respect to a normalized charge $e$
for extreme black holes.
The Kerr-Newman black hole has the maximum value of
rotation parameter, which is $q=1$.
On the other hand, The maximum value for a CBMPV black hole
can be larger than unity if the size is small ($r_g<(32/27) R_5$).}
\label{fig3}
\end{figure}

For a CBMPV black hole, we have a constraint of $\lambda^2\leq 1/q^2$.
Suppose two charges are the same ($Q_2=Q_5$). Writing $\lambda^2$ in terms of
$e_2=e_5\equiv e$, we find a relation between $q$ and $e$ in the extreme
limit (the equality of the above condition) as
\bea
q^2=e^2(2-e){R_5\over r_g}
\ena
which is shown in Fig.~\ref{fig3}.
The maximum value of $q$ is $q_{\rm max}^2=32/27\times
(r_g/R_5)^{-1}$ at $e=4/3$.
Hence if $r_g/R_5<32/27$, the value of $q$ at
 maximum rotation exceeds unity (the Kerr bound).
This fact may be important for the following reason.
In four-dimensional general relativity,
rotation of a black hole is bounded by
a critical value (the Kerr bound [$q=1]$), beyond which
there appears a naked singularity.
However, if we discuss a black hole in superstring/M-theory,
we can find an effectively four-dimensional
compact object with $q>1$ without a naked singularity.
Observation of such a regular compact object with  $q>1$
might provide us an indirect evidence of extra dimensions.

A similar discussion gives a bound on the size of black hole.
The constraint $\lambda^2\leq 1/q^2$ imposes that
the size of a black hole must be smaller than some critical value,
i.e. $r_g\leq r_g^{({\rm cr})}$, where
\bea
r_g^{({\rm cr})}=2G_4M^{({\rm cr})}\equiv
{32\alpha^{(2)}\alpha^{(5)}\over   (1+\alpha^{(2)}+\alpha^{(5)})^3
q^2}~ R_5 \,.
\ena

Let us check how restrictive this constraint is.
If the black hole is rapidly rotating, i.e. $q\sim O(1)$ and all charges
are of the same order of magnitude,
i.e. $\alpha^{(2)}\sim \alpha^{(5)}\sim O(1)$,
then we find $r_g^{({\rm cr})} \sim  R_5$.
As for the  scale of extra dimension $R_5$, we have a constraint by
the experiment of Newtonian gravity, i.e.
$R_5 <$ 0.1 mm~\cite{mm_gravity}, which gives the upper limit
for the mass of CBMPV black hole, that is, about $10^{27}$ g.
A rapidly rotating CBMPV black hole could be realized
if its mass is smaller than this critical value.

\subsection{supersymmetry}

The asymptotic metric can be considered to describe effectively
a four-dimensional rotating object.
It contains only one rotation, that is ${\cal A}_{\phi}$.
No ${\cal A}_{\psi}$ appears in the effective four-dimensional metric.
However, supersymmetry is preserved if and only if
${\cal F}_{ij}$ is self-dual, i.e. $J_\phi=-J_\psi$.
Since the effective metric does not contain the information
about $J_\psi$, one may wonder whether this spacetime is supersymmetric or not.
In particular, non-supersymmetric CBMPV black hole ($J_\phi\neq -J_\psi$)
also gives the same asymptotic structure as that of a supersymmetric one
if it has the same value of $J_\phi$.

To understand this situation, we should not forget the fifth direction.
Although the effective spacetime looks four-dimensional, there is a small
circle $S^1$ with the radius $R_5$ in the fifth direction.
In the Kaluza-Klein theory, when we compactify the five-dimensional spacetime on
four-dimensional one, we find not only four-dimensional gravity but also
an electromagnetic field and a scalar field coming from the metric in the
compactified dimension.
When we consider the far region in the present solution,
the four-dimensional metric does not contain ${\cal A}_\psi$, but
it appears as the ``electromagnetic" field.
So, when we analyze the asymptotic structure of this spacetime,
we have to discuss the Einstein-Maxwell system.
We can check that supersymmetry is preserved
if and only if the magnitude of rotation (${\cal A}_\phi$)
balances with that of the ``electromagnetic" field (${\cal A}_\psi$).

\section{Concluding Remarks}

In this paper we have presented a supersymmetric rotating solution
with asymptotically flat four-dimensional spacetime times a circle.
It is constructed by superposing infinite number of
BMPV black holes aligned in the fifth direction
in five dimensions. This is possible because the solution is supersymmetric
and so the field equation governing the solution is linear.
We have examined its properties including mass, area and entropy.
Considering the five-dimensional solution from the asymptotical
four-dimensional point of view, we have also shown that the rotation parameter of
maximally rotating black hole exceeds unity (the Kerr bound) if
the size of black hole is small ($r_g\leq (32/27) R_5$).

It would be interesting to compare our solution with
another supersymmetric  rotating black hole solution with
asymptotically flat four-dimensional spacetime (the RING black hole).
However, because those solutions contain different type of charges,
such a comparison may not make sense, even if we fix four-dimensional observables
such as the ADM mass $M$, angular momentum $J$ (or a rotation
parameter $q=J/G_4M^2$) of a black hole, and a compactification radius $R_5$.

So far, not so many supersymmetric rotating solutions with asymptotically
flat four-dimensional spacetime (+ small extra dimension) are known.
It would be interesting to find further examples of
effective four-dimensional
rotating supersymmetric solutions
 and study their properties.

\acknowledgments

We would like to thank Roberto Emparan, Gary Gibbons,
and Harvey Reall for valuable comments.
This work was partially supported by the Grant-in-Aid for Scientific Research
Fund of the JSPS (Nos. 16540250 and 17540268) and for the
Japan-U.K. Research Cooperative Program, and by the Waseda University
Grants for Special Research Projects and for The 21st Century
COE Program (Holistic Research and Education Center for Physics
Self-organization Systems) at Waseda University.
KM would like to thank DAMTP, the Centre for Theoretical Cosmology, and Trinity
College for hospitality during his stay.
He would also acknowledge hospitality of Gravitation and Cosmology Group
at University of Barcelona, where this work was completed.


\end{document}